# Driving Safety Prediction and Safe Route Mapping Using In-vehicle and Roadside Data

Yufei Huang, Mohsen Jafari, and Peter Jin

*Abstract*— Risk assessment of roadways is commonly practiced based on historical crash data. Information on driver behaviors and real-time traffic situations is sometimes missing. In this paper, the Safe Route Mapping (SRM) model, a methodology for developing dynamic risk heat maps of roadways, is extended to consider driver behaviors when making prediction. An Android App is designed to gather drivers' information and upload it to a server. On the server, facial recognition extracts drivers' data, such as facial landmarks, gaze directions, and emotions. The driver's drowsiness and distraction are detected, and driving performance is evaluated. Meanwhile, dynamic traffic information is captured by a roadside camera and uploaded to the same server. A longitudinal-scanline-based arterial traffic video analytics is applied to recognize vehicles from the video to build speed and trajectory profiles. Based on these data, a LightGBM model is introduced to predict conflict indices for drivers in the next one or two seconds. Then, multiple data sources, including historical crash counts and predicted traffic conflict indicators, are combined using a Fuzzy logic model to calculate risk scores for road segments. The proposed SRM model is illustrated using data collected from an actual traffic intersection and a driving simulation platform. The prediction results show that the model is accurate, and the added driver behavior features will improve the model's performance. Finally, risk heat maps are generated for visualization purposes. The authorities can use the dynamic heat map to designate safe corridors and dispatch law enforcement and drivers for early warning and trip planning.

## I. Introduction

Motor vehicle crashes have a tremendous economic and societal impact worldwide. Measuring roadway safety plays a significant role in reducing the risk of road traffic injuries and death. Risk measurement of roadways is commonly practiced based on historical crash data. The risk measures computed using data from past circumstances cannot correctly reflect the road safety issues caused by fast-changing events such as heavy traffic and weather conditions [1]. Also, the lack of pre-crash data makes it hard to study the conflict probability, predict the conflicts, and send a warning in advance [2]. Moreover, more than 90% of vehicle crashes are caused by human errors, such as speeding, fatigue, drunk, and distracted driving [3]. However, many traffic data analytics platforms, such as RITIS and TOMTOM, hardly consider drivers' behavior in an accident because of lacking in-vehicle information.

Taking advantage of the fast-developing communication and Vehicle-to-Infrastructure (V2I) technologies, dynamic traffic conflict data and vehicle interior information can be retrieved and analyzed. The performance of wireless technologies, such as dedicated short-range communication (DSRC), Wi-Fi, and 5G, has been evaluated and proved capable of hosting communications between vehicles and roadside infrastructure [4]. Using roadside infrastructures, such as cameras and LiDAR, vehicle detection and tracking generate vehicle speed and trajectory profiles efficiently [5]. Also, driver behaviors can be collected using cameras placed inside a vehicle and then uploaded to roadside servers for further analysis.

There are significant gaps between theoretical advances and real-world practices in analyzing driving safety: (i) The driving performance measurement lacks driver behavior data because the information is hard to capture in real-time or requires additional devices inside the vehicle. (ii) Roadway safety prediction models seldom consider the status of individual drivers, such as their emotions and attention. (iii) Feature engineering steps for road safety analysis are required to examine multiple inputs, including driver behavior, roadway characteristics, vehicle speed profile, and vehicle status. A proper model must be established to combine different types of inputs to enhance the prediction performance. In this work, the authors set up a roadside camera to get a bird's eye view of an intersection in real time and develop an Android application to capture the video of the driver's face. Both videos are uploaded to a server hosted by Amazon Web Services (AWS) via a 5G network. Vehicle trajectory information and driver behaviors are detected from the videos and combined based on the timestamp that the vehicle enters the intersection. Then, this paper extends the Safe Route Mapping (SRM) methodology developed in [6] to calculate the risk level of roadways. The SRM is a data-driven and driver-based method that combines over-a-trip safety data and historical crashes to predict risk scores and create dynamic risk heat maps of roadways. The risk score model contains multiple data sources, including historical crash counts, driver behaviors, road characteristics, and vehicle status. In principle, this paper's contributions are three folds: (i) A way to collect traffic data from roadside and driver's information inside a vehicle simultaneously. The proposed

Y. Huang, and M. Jafari are with the Department of Industrial and Systems Engineering, Rutgers University – New Brunswick, Piscataway, NJ 08854, USA (e-mail: yh639@scarletmail.rutgers.edu; jafari@soe.rutgers.edu).

P. Jin is with the Department of Civil Engineering, Rutgers University – New Brunswick, Piscataway, NJ 08854, USA (e-mail: peter.j.jin@rutgers.edu).

method introduced a way to manage driver behaviors when analyzing road safety. (ii) An improved SRM model that considers human factors. The model can predict the safety scores of road segments in the next one or two seconds. (iii) The safety scores can produce risk heat maps that help transportation authorities designate safe corridors and dispatch law enforcement. Individual users can also benefit from being alerted about the predicted high-risk area during a trip.

This paper is organized as follows. Section II presents an overview of road-safe measurement methods and computer vision techniques for vehicle trajectory recognition. Face recognition and driving performance evaluation methods are also reviewed in this section. Section III describes the data collection process and explains the longitudinal-scan-line method to obtain vehicle trajectory data. Section IV illustrates our Light Gradient Boosting Machine (LightGBM) and Fuzzy logic models to calculate risk scores. Section V shows the model training process and the validation results of our proposed method. Section VI presents a simulation platform to gather more data and validate the improved SRM model. Section VII is the summary and provides some future thoughts.

## II. Literature Review

The Highway Safety Manual (HSM) published by the American Association of State Highway and Transportation Officials (AASHTO) outlines that developing Safety Performance Functions (SPFs) can be used by jurisdictions to make better safety decisions [7]. The current practice of using SPF to predict average crash frequency suffers from insufficient data and model accuracy [8]. Considering vehicle crashes as rare events compared to a large amount of daily traffic, traffic conflicts or near misses are a rough approximation for studying traffic accidents [9]. Near-miss, an accident barely avoided, is identified using specific Traffic Conflict Indicators (TCIs) [10]. Common TCIs that people use are Time-to-Collision (TTC), Modified Time-to-Collision (MTTC), Deceleration Rate to Avoid a Crash (DRAC), and Post Encroachment Time (PET). Considering human error counts for approximately 87% of all commercial vehicle crashes, naturalistic driving studies (NDS) intended to observe and record driver behavior in real-time [11]. Driver-based data have been used to analyze collisions and crash surrogates and develop collision avoidance advisory systems [12]. The 100-Car Naturalistic Driving Study sponsored by the National Highway Traffic Safety Administration (NHTSA) and the Virginia Department of Transportation (VDOT) is the first instrumented vehicle study to collect large-scale naturalistic driving data [13]. However, the driver-based and vehicle-based data approaches still suffer from a non-reliable or low-density level of data [6]. In recent years, with the development of V2I technology and the lower cost of infrastructure sensing, collecting data inside a vehicle and from roadside infrastructure has become possible.

Inside a vehicle, a driver concentration control system that monitors the driver's behavior mainly uses cameras, eye-trackers, or contactless sensors. [14] provides a solution that driver's performance can be captured using a smartphone application, which is more affordable than adding additional devices inside a vehicle. Videos of the driver are streamed to the server, where the facial recognition system takes place. Deep learning has recently enjoyed fame in face detection, face alignment, and feature extraction. Trained with convolutional neural network (CNN) [15], LFPW [16], and Helen [17] training sets, facial behavior analysis toolkits such as OpenFace can monitor facial landmark motion, head pose (orientation and motion), and eye gaze [18]. [19] proves that the driver's emotion affects driving performance, which can be modeled by a U-shaped relationship between performance, arousal, and valence. Given the facial landmark motion, researchers have figured out a way to identify facial expressions with the Facial Action Coding System (FACS) [20]. Also, based on the facial landmarks, a drowsiness alarm can be designed according to the eye aspect ratio (EAR) [21]. Head pose and eye gaze indicate whether the driver is focusing on the road.

On the roadside, computer vision sensors based on traffic surveillance systems and cloud computing provides new opportunities for enhancing traffic safety. Vehicle detection and tracking are the two main steps of traffic video analysis. The detection separates the vehicle from the background, and the tracking step traces object movements. Faster Region Proposal Convolutional Neural Network (R-CNN) [22], Single Shot MultiBox Detector (SSD) [23], and You Only Look Once (YOLO) [24] are commonly used model-based methods to localize and classify vehicles based on their shapes. These methods require large training datasets, complicated deep neural network structures, and a long training time to achieve optimal performance [25]. [26] introduced a computer vision algorithm to obtain vehicle trajectories from high-angle traffic video. Their model combines scanline-based trajectory extraction and feature-matching coordinate transformation to detect vehicles and get their traces. Their method has been proven to be robust and accurate.

The analysis of NDS reveals correlations between driver behaviors and roadway segments in either crash/near or regular situations. [27] provides a data-driven approach for drivers' safety risk profiling and roadway segments' safety risk scoring. In their work, elastic net regularized multinomial logistic regression is applied to enhance the prediction performance of the road safety model. [6] further develops the model using neural networks and creating risk heat maps for visualization purposes. In their work, driver behaviors are limited to actions like speeding, stop sign violations, and rapid lane change, which is not comprehensive enough.

## III. Data Collection

In this section, the data collection and pre-processing steps shall be discussed. The section introduced the architecture of the data collection process that retrieves the videos from the roadside infrastructure and inside the vehicles. Also, the applied vehicle trajectory recognition and drivers' behavior detection methods are presented.

*A. Data flow diagram*

Fig. 1 shows the data flow diagram of the data collection process. Firstly, raw data collected by the roadside camera and Android APP are uploaded to the Cloud Server via the 5G network. The vehicle profile created by vehicle trajectory recognition and the driver behavior data created by the driver's face recognition is synchronized and merged based on the GPS information (Longitude and Latitude) and timestamp. For example, suppose the Android APP reports the vehicle at specific GPS coordinates that are also detected by the roadside camera simultaneously. In that case, we then conclude they represent the exact vehicle. Next, python programs generate vehicle trajectory and driver behavior profiles using the uploaded raw data in real-time. Finally, risk scores for each road segment are predicted using the safety risk model. The risk heat maps are created based on the risk scores and are transmitted to the Android APP to warn drivers about the traffic conflict probabilities. The safety risk model and the risk heat map will be discussed in Section IV. AWS is selected to host the database and perform cloud computing in this work.

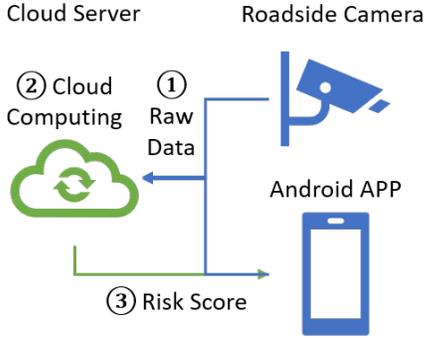

Fig. 1. The dataflow diagram of the data collection process.

### B. Vehicle trajectory recognition

Videos of the roadway are filmed by a camera placed on the roadside infrastructure and uploaded to the cloud server in real time. On the server, a scanline-based trajectory extraction technique is applied to recognize the vehicles and obtain each vehicle's trajectory in the videos. The method was first brought up in [26], and it separates vehicle strands from pavement background on the spatial-temporal diagram by considering color features, gradient features, and motion features [28].

As is shown in Fig.2, the first step is to manually label the traffic lanes (scanlines) in the video. Then, a spatial-temporal (ST) map is generated using frame-by-frame stacking pixel values on each scanline. The ST map represents the time progression of pixels from moving vehicles on the lane. Each vehicle passing through the scanline will leave a group of strands that can be used to generate a pixel trajectory of a car. The main challenge during this step is to suppress noise. The prominent noises of the ST Map are the static objects (traffic signs and road signs) and the shadows of vehicles from adjacent lanes. The detailed algorithm for identifying and removing noises can be found in [26].

After the ST map is generated, we can find the vehicle pixel coordinates in the video coordinates $(u, v)$ on each frame. $u$ represents the horizontal pixel, and $v$ is the vertical pixel. The second step is transforming the video coordinates into real-world coordinates based on $(x, y, z)$. $x$ is the longitude value, $y$ is the latitude value, and z is the vehicle's altitude at each timestamp in the video. Feature matching is applied to detect trajectory points in the ST map to their real-world coordinates. As is shown in Fig. 2, the actual location of a point in the video can be estimated in Google Maps.

Following the above steps, the vehicle trajectory profiles are created. The vehicle's front bumper is selected to represent the vehicle's position on each frame. Given that the video is recorded at 20 frames per second and their real-world coordinates, we can calculate the speed and acceleration information for each vehicle at each timestamp. In the data pre-processing process, more features are generated based on the raw data from the roadside camera. These inputs can be classified into four categories: a) Roadway characteristic - The shape of the lane that the vehicle is on, such as going straight, turning left, turning right, merging, and diverging. b) Vehicle status - Including information such as whether the vehicle is leading a queue or following other vehicles and the angle between its direction and the front car. c) Speed profile - Such as speed, acceleration, and the speed difference to the front car. d. distance profile. Such as the headway and safety distance according to vehicle type [6]. The created features will be input

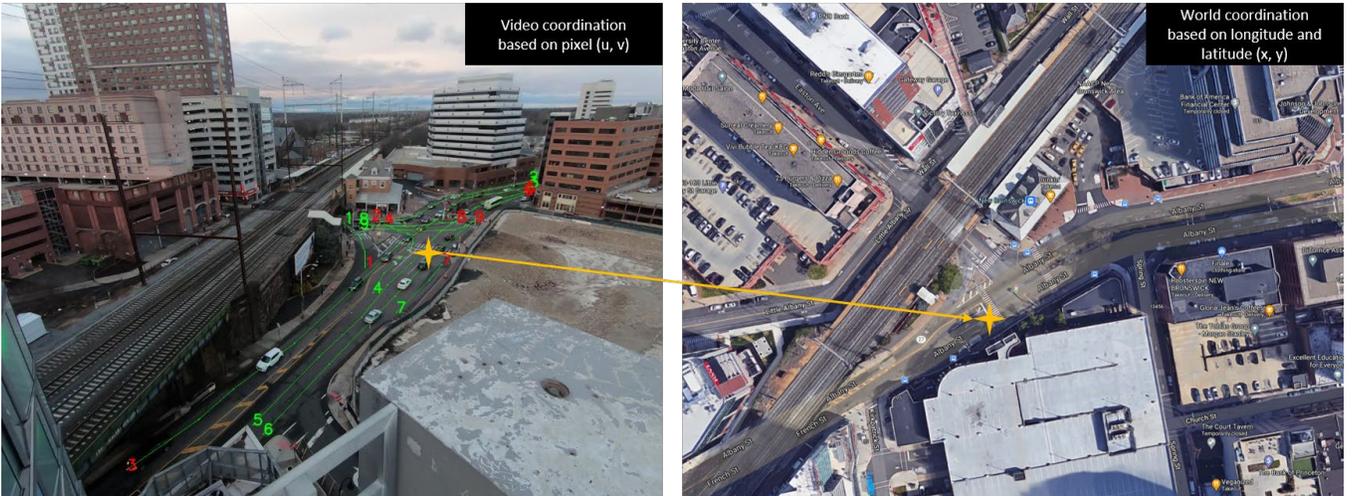

Fig. 2. Transformation of Coordination.

variables in our traffic conflict prediction model, which will be discussed in section IV.

*C. Driver behavior analysis*

An Android application is developed to collect drivers' behavior data inside a vehicle. Drivers can place their Android phones on the dashboard or an air vent in a phone holder. The front camera needs to aim at the face of the driver. As is shown in Fig. 3, the APP is designed to capture the video of the driver's face and then upload the video to the same cloud server. Vehicle speed, acceleration, and GPS information will also be recorded and uploaded.

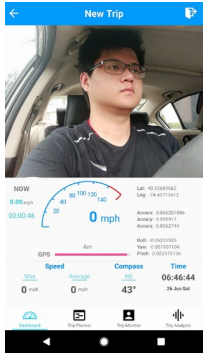
Fig. 3. The user interface of the Android application.

The driver's video is processed on the server using the OpenFace toolkit to recognize the face and extract the facial landmarks. OpenFace is an open-source facial behavior analysis toolkit capable of detecting facial landmarks, estimating head pose and eye gaze, and recognizing facial action units (AU) [29].

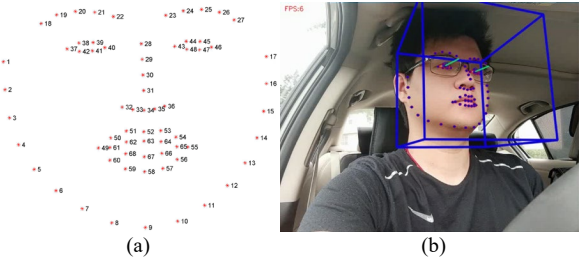
Fig. 4. Facial recognition based on OpenFace toolkit.

OpenFace utilizes deep neural networks to represent the face on a 128-dimensional unit hypersphere. It can detect 68 facial landmarks, as shown in Fig. 4 (a). These landmarks are critical points on a human face that localize and label the facial regions like the mouth, eyebrows, eyes, nose, and jaw. As is shown in Fig.4 (b), OpenFace not only estimates the coordinates of the face in the image but also provides the angles of head pose and eye gaze. Assume $x_h$ and $y_h$ as the angle of head pose on $x$ and $y$ axis. Let $x_e$ and $y_e$ be the angle of eye-gaze on $x$ and $y$ axis. Using (3), we can calculate the angle of the driver's focusing area represented by the angle on x axis $x_f$, and the angle on y axis $y_f$. $x_f$ is positive when the driver is looking to the left, and it is negative when the driver is looking to the right. $y_f$ is positive when the driver is looking up, and it is negative when the driver is looking down.

$$\begin{cases} x_f = x_h + x_e \\ y_f = y_h + y_e \end{cases} \quad (3)$$

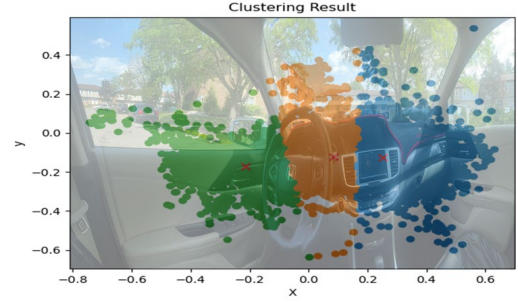
Fig. 5. The approximation of driver's focusing area.

K-Nearest Neighbors algorithm (K-NN) [30] is applied to cluster the points that represent where the driver is looking using the focusing area coordinates $(x_f, y_f)$. As is shown in Fig. 5, the pre-collected focusing points are grouped into 3 clusters. Based on where each point is locating at, we can approximately understand whether the driver is looking straight ahead or not.

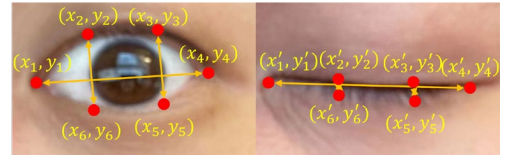
Fig. 6. Facial landmarks around the eyes.

EAR is commonly used for fatigue level classification. If the eye-blinking frequency radically increases, we would observe continuous changes in EAR [31]. EAR ($r_e$) can be calculated using (4) given the coordinates of the facial landmarks around the eyes, as shown in Fig.6.

$$r_e = \frac{\sqrt{(x_2-x_6)^2 + (y_2-y_6)^2} + \sqrt{(x_3-x_5)^2 + (y_3-y_5)^2}}{2 \cdot \sqrt{(x_4-x_1)^2 + (y_4-y_1)^2}} \quad (4)$$

The threshold of $r_e$ is set at 0.26 based on an experiment shown in Fig. 7. The eyes are considered closed when $r_e$ is smaller than the threshold.

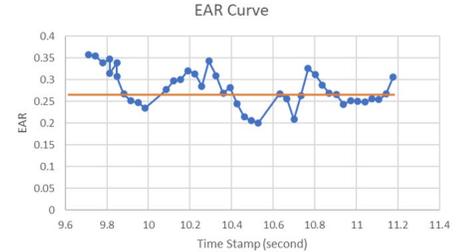
Fig. 7. The approximation of driver's focusing area.

Similarly, the mouth aspect ratio (MAR) is created to indicate whether the driver's mouth is opened or not. The threshold of MAR is set as 0.05 based on experiments. If MAR ($r_m$) goes beyond the threshold, the driver might be talking or eating, which distracts the driver from focusing on the driving task.

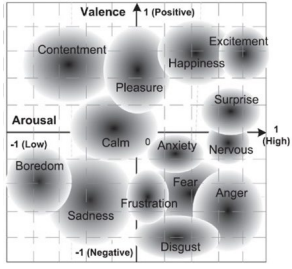

Fig. 8. The Arousal-Valence Plane [19].

Besides drivers' focusing area and fatigue level, emotion also affects human performance through influencing individuals' judgment and behavior. For example, stressed operators could not achieve optimal performance in complex task environments [32]. The Facial Action Coding System (FACS) determines the driver's emotion. FACS refers to a set of facial muscle movements that correspond to a displayed emotion. The basic elements of FACS are called action units (AU) [20], which include 46 main action units, such as inner brow raiser, outer brow raiser, cheek raiser, and nose wrinkle; 8 head movement action units, such as head turn left, head turn right, head up and head down; and four eye movement, including eyes, turn left, right, up, and down. Specific joint activities of facial muscles pertain to a displayed emotion. For example, happiness is determined by the combination of action units 6 (cheek raiser) and 12 (lip corner puller). Sadness is identified by unit 1 (inner brow raiser), unit 4 (brow lowered), and unit 15 (lip corner depressor). Action units can be determined by the movement of facial landmarks based on their definition. In this work, we only consider six basic emotions: happiness, sadness, surprise, fear, anger, and calm. [19] proposed a way to quantify emotion by a Valence-Arousal (VA) plane, as is shown in Fig. 8. The proposed 2-dimensional emotion model can represent emotion combinations rather than pure emotions. The coordinates of each emotion on the VA plane can be found in Table I.

TABLE I
COORDINATES OF EACH EMOTION ON THE VA PLANE [19]

| Emotion | Valence | Valence Range | Arousal | Arousal Range |
|---|---|---|---|---|
| Happiness | 0.23 | (0.08, 0.38) | 0.31 | (0.18, 0.44) |
| Sadness | -0.18 | (-0.36, 0) | -0.26 | (-0.44, -0.08) |
| Surprise | 0.37 | (0.25, 0.48) | 0.08 | (-0.01, 0.17) |
| Fear | 0.21 | (0.09, 0.33) | -0.24 | (-0.36, -0.11) |
| Anger | 0.36 | (0.21, 0.5) | -0.3 | (-0.46, -0.15) |
| Calm | -0.09 | (-0.26, 0.08) | -0.11 | (-0.24, 0.02) |

Yerkes-Dodson law points out that performance increases with physiological or mental arousal, but only up to a point. When levels of arousal become too high, performance decreases [33]. The process is often illustrated graphically as a bell-shaped curve that increases and then decreases with higher arousal levels, as shown in Fig. 9. A bivariate normal distribution can model this two-dimensional emotion-performance relation.

Experiments that use different ways of measuring driving performance show that we can get other spots to achieve the best performance (sweet point) [19]. If we define driving performance as the number of traffic violations, the coordinate of the sweet point on a VA plane is $(0, 0.2)$. If we take lane deviation as the judgment of driving performance, the sweet point locates at $(0, 0.1)$. If the judgment is based on brake reaction time, the sweet point is $(0.2, 0.2)$. All three sweet points fall in the VA plane's calm area, indicating that the driver can achieve the best performance when the driver is calm. We are going to add the probability density functions (PDF) of all three bivariate normally distributed models up as the driver's performance score. Each model is generated by one of the three sweet points.

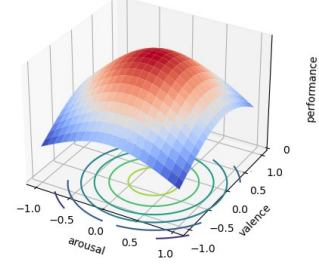

Fig. 9. The Arousal-Valence Plane.

In summary, we have created four more features based on the driver's behavior: focusing area, EAR, MAR, and performance score based on emotion. These features will also be used as inputs in the traffic conflict prediction model, which will be discussed in section IV.

IV. IMPROVED SRM MODEL

This section introduces how to use the features created in the previous section to predict the traffic conflict indicators (TCIs) for an individual vehicle. These TCIs are used to calculate the risk scores for road segments. The risk scores can be used to generate risk heat maps, which show the changes in risk levels of each road segment over time.

A. Traffic conflict prediction

TCIs are developed to study near misses, which occur more frequently than collisions. Analyzing TCIs can help notify drivers to take action before collisions happen. However, research has shown that adopting different indicators will give us different labels of conflicts [34]. Therefore, we combine the results provided by various indicators in this study. We use three commonly used TCIs, which are: (i) time to collision (TTC), (ii) modified time to collision (MTTC), and (iii) deceleration rate to avoid a crash (DRAC). TTC ($I_T$) estimates the time to a crash based on current distance and speed and is defined in (5). $x_L$ and $v_L$ represent the longitudinal coordinate and speed of the leading vehicle. $x_F$ and $v_F$ represent the longitudinal coordinate and speed of the following vehicle. The commonly set threshold value for TTC is 1.5 seconds. TTC smaller than 1.5 seconds implies a near miss. MTTC ($I_M$) improves TTC by considering acceleration and is defined in (6). $\Delta v$ means the speed difference between the leading and following vehicles. $\Delta a$ means the acceleration difference between the two cars. The commonly used threshold for MTTC is also 1.5 seconds. A near miss happens when MTTC is less than 1.5 seconds. Finally, DRAC ($I_D$) shows how much deceleration the following vehicle needs to avoid crashing into the front car. It is defined in (7). The threshold value for DRAC is $3.35 m/s^2$. It is safe when DRAC is smaller than the threshold.

$$I_T = \frac{x_L - x_F}{v_L - v_F} \quad (5)$$

$$I_M = \frac{\Delta v \pm \sqrt{\Delta v^2 + 2\Delta a(x_L - x_F)}}{\Delta a} \quad (6)$$

$$I_D = \frac{(v_F - v_L)^2}{2 * (x_L - x_F)} \quad (7)$$

Light Gradient Boosting Machine (LGBM) [35] was introduced to predict whether TTC, MTTC, or DRAC go beyond the threshold in the next 1 second or 2 seconds. LGBM is a gradient boosting framework that uses tree-based learning algorithms. Instead of building an ensemble of deep independent trees (Random Forest), LGBM creates an ensemble of shallow and weak successive trees. Each tree learns and improves the previous results. LGBM has a faster training speed and accuracy than any other boosting algorithm. Fig. 10 illustrates the LGBM structure used in this work. The inputs can be classified into six categories:

(i) "Driver Behavior" includes the features we create in section III-C. In addition, "Type 1" and "Type 2" are dummy variables and are used to tell whether the driver is average (type 1) or aggressive (type 2). If a driver drives 8 miles per hour beyond the speed limit, or the driver's maximum acceleration rate is beyond $5\ m/s^2$, or the driver's maximum deceleration rate is beyond $3\ m/s^2$, we will consider the driver is type 2. Otherwise, the driver is type 1 [36].

(ii) "Road Characteristic" describes the shape of the lane that the driver is on. They are dummy variables that indicate whether the lane that the driver is driving on is going to diverge, merge, turn left or right, etc.

(iii) "Vehicle Status" shows whether the vehicle is leading or following another car. "In Queue" is a binary variable. Direction means the yaw angle of the car. The angle to the front vehicle shows the angle between our vehicle's heading direction and the direction to the front vehicle. These are inputs we created from section III-B.

(iv) "Distance" includes headway distance (HDWY), which is the distance between two successive vehicles' front bumpers. The safe distance is related to the type of driver. When the driver changes the lane, a safety distance reduction factor will be applied to their original car-following space [37].

(v) "Speed" contains the vehicle's current speed and acceleration. It also indicates the speed difference between the front car and the host vehicle. These are inputs we created from section III-B.

(vi) The last category contains TTC, MTTC, or DRAC values in the previous one second and the current time. For example, when we want to predict TTC in the next one or two seconds, we put the TTC value in the previous one and at the current time point here. When predicting MTTC or DRAC, we replace these inputs with the corresponding values. Therefore, we use 6 LGBM models with a similar structure for predicting these Indicators.

The output of each LGBM model is a binary variable that indicates whether one of the three TCIs will go beyond the threshold in the next one second or two seconds. The outputs can take values of either 0 (no conflict) or 1 (conflict).

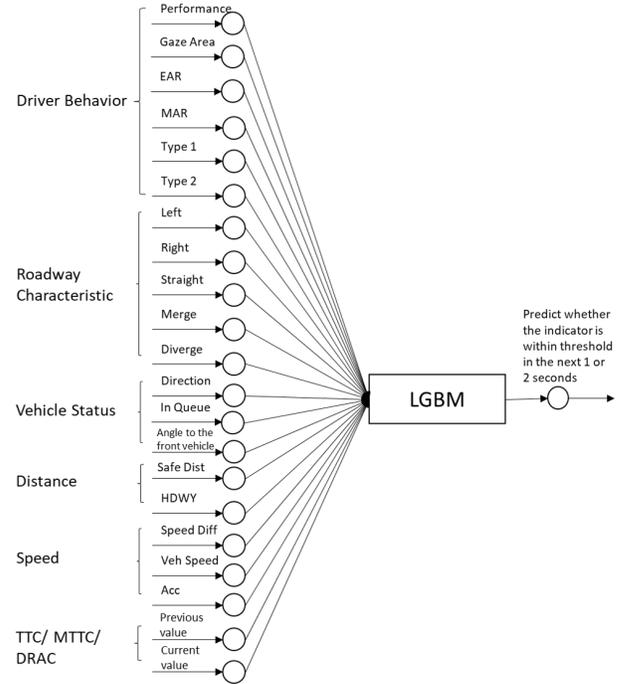

Fig. 10. Structure of the LGBM.

Note that not all vehicles have in-vehicle data. For the drivers who cannot or are not willing to use the Android APP to share their information, a simplified LGBM model is trained and used. The simplified version removes the inputs in the category "Driver Behavior" except "Type 1" and "Type 2". These two inputs can still be observed using roadside video.

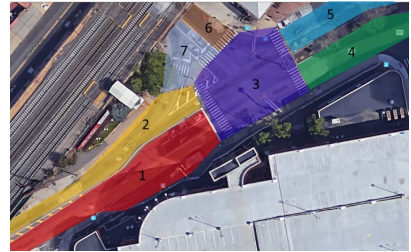

Fig. 11. Illustration of road segments.

### B. Risk score

After the TCIs are predicted at each timestamp, we will calculate risk scores for each vehicle. Fig. 11 provides an example of how road segments are divided. The intersection is divided into seven parts based on the direction of the lane.

Besides the real-time TCI predictions, historical crash data are also included in our risk score model. A Fuzzy Logic model combines the average crash counts per year of the road segment where the vehicle is located and the prediction of three TCIs at each time point. The structure of the Fuzzy Logic model is shown in Fig. 12. TCI predictions can take values of either 0 (no conflict) or 1 (conflict). Average crash counts per year are divided into three levels: (1) low: 0-5 crashes per year, (2) medium: 6-10 crashes per year, and (3) high: over ten crashes per year. In this case, we established 24 rules described in Table II for the Fuzzy Logic model to calculate the risk score on a 0 to 100 scale. The risk score of the road segment is defined by the maximum risk score of the vehicle in that segment at each time point.

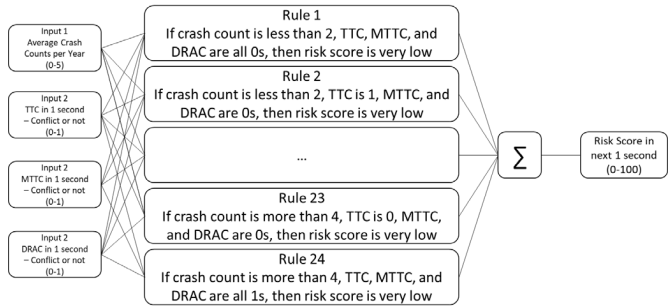

Fig. 12. Structure of the Fuzzy Logic model.

TABLE II
RULES FOR THE FUZZY LOGIC MODEL

| No. | Rules | Risk level |
|---|---|---|
| 1-8 | Number of historical crashes is large or medium and at least two TCIs predict a conflict | Large |
| 9-18 | Number of historical crashes is large or medium and at most one TCI predicts a conflict; Number of historical crashes is small and at least two TCIs predict a conflict | Medium |
| 19-24 | At most one TCI predict a conflict | Small |

*C. Risk heat map*

The risk heat map is created to visualize the risk scores for each segment. The risk scores are divided into five very small levels (if the risk score falls in [0,20)), small (if the risk score falls in [20,40)), medium (if the risk score falls in [40,60)), large (if the risk score falls in [60,80)), and very large (if the risk score falls in [80,100]). The corresponding colors assigned to each level are green, blue, yellow, orange, and red. Based on these rules, a cell plot is created to show the value of risk scores and the predicted risk scores for each segment at each time point. Fig. 13 shows an example of the risk heat map. We predict the risk scores for each road segment at every time point. Then, the actual risk scores are calculated based on the factual data collected in the next 1 or 2 seconds to validate the prediction results. In the plot, each row corresponds to a road segment in Fig. 11. The first and third columns show the actual risk score in the next 1 or 2 seconds, and the second and fourth columns are our predicted results for the risk score in the next 1 or 2 seconds. By taking an average of the risk scores of each road segment in specific time windows, Fig. 14 illustrates the changes in road risks through time. Also, Google Maps is customized to show the prediction results using Google Map API, as shown in Fig. 15. The personalized map can provide a clear view of the safety situation on the road, and the map will be sent back to the Android APP. The APP users can be alerted about the predicted high-risk area during a trip.

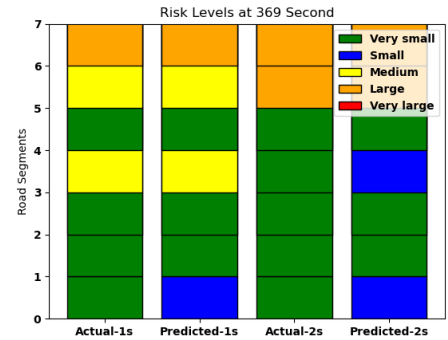

Fig. 13. An example of the risk heat map at a certain timestamp.

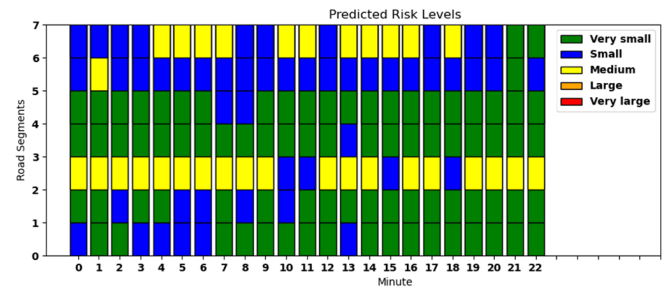

Fig. 14. Risk heat map through time.

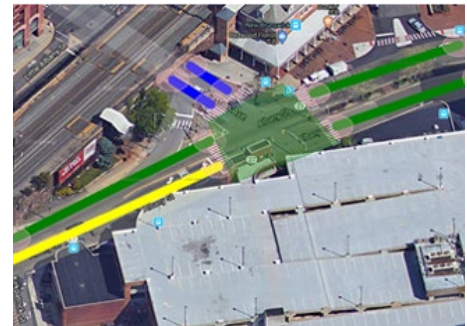

Fig. 15. Customized Google map.

V. MODEL TRAINING AND VALIDATION

This section demonstrates how the SRM model is trained using actual traffic data collected from an intersection. The validation results are also presented.

*A. Scope of the testbed*

The testbed locates at the intersection of Easton Avenue and Albany Street in New Brunswick, New Jersey, USA. The roadside camera is placed on the roof of a public garage. It provides a bird's eye view of the intersection, as shown in Fig. 2. The New Brunswick train station located at the intersection makes the traffic condition more complicated.

Four test drivers were invited to join the experiment and used the Android APP to collect their facial video during the test, as is shown in Fig. 16. The experiment was performed at 3:35 PM on May 16[th], 2021, and it lasted 40 minutes. The data collected from the experiment was divided into two parts. In the first 20 minutes, the collected roadside video and drivers' information are used to train the LGBM model mentioned in Section IV-A. In the second half, we evaluate the accuracy of the LGBM model and the fuzzy logic reasoning model using the rest data. There are 1,833 recognized vehicle trajectories, including the exact vehicle with multiple trips passing through the area. The test drivers were asked to keep driving around and driving through the testbed as often as possible. Among all the recognized trajectories, 36 trajectories were made by the four test drivers. Their driving behavior information was matched with their speed profiles collected by roadside video. For the other trajectories that miss the driver's information, a simplified LGBM model, which has no driver behavior inputs, was trained separately, as discussed in Section IV-A.

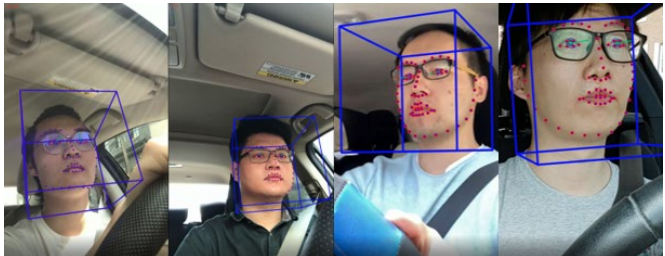

Fig. 16. Four test drivers for the experiment.

### B. Validation results

One limitation of OpenFace is that the facial recognition result is not stable in poor illumination conditions. The driver's face can be detected in low light conditions, as shown in Fig. 17 (a). The testbed has enough street lights at night, and the model works. However, the driver's face cannot be seen when it is too dark, as shown in Fig. 17 (b). When street light is not bright enough, the authors suggest removing driver behavior features from the proposed LGBM model and using the simplified model to predict the safety index in dark environments.

Notice that 16.2% of the training data detect a conflict based on either TTC, MTTC, or DRAC, indicating that the classification's training set is imbalanced. Synthetic Minority Oversampling Technique (SMOTE) is utilized to deal with the problem. SMOTE handles an imbalanced classification in which there are too few examples of the minority class for a model to learn the decision boundary effectively [38]. The basic idea of SMOTE is to select samples close to the feature space, draw a line between the data points in the feature space, and create new samples along that line. After applying SMOTE, the number of near misses and non-conflict cases becomes the same. 10-folds cross-validation is also applied when training the LGBM models to avoid over-fitting. Two test cases are created to elaborate on the influence of SMOTE: (i) train models without applying SMOTE on the training dataset, and (ii) train models after applying SMOTE on the training dataset. The result shows that all models perform well even without SMOTE. Applying SMOTE to the training data can improve the prediction accuracy for all models.

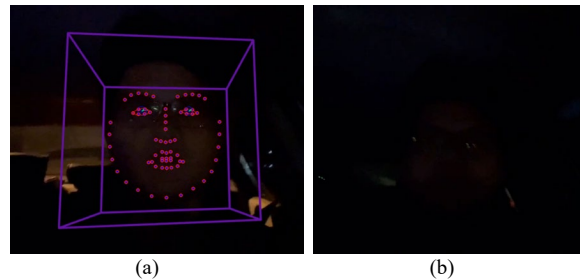

(a) (b)

Fig. 17. Facial recognition in bad illumination conditions.

TABLE III
PREDICITON ACCURACY WITH OR WITHOUT SMOTE

| Case | TTC in next 1s | TTC in next 2s | MTTC in next 1s | MTTC in next 2s | DRAC in next 1s | DRAC in next 2s |
|---|---|---|---|---|---|---|
| i | 0.978 | 0.979 | 0.961 | 0.937 | 0.944 | 0.928 |
| ii | 0.999 | 0.998 | 0.990 | 0.974 | 0.978 | 0.972 |

The feature importance of the LGBM models is shown in Fig. 18. The speed difference to the front vehicle, headway distance, Vehicle speed, acceleration, and the angle to the front vehicle play the most important roles when predicting whether there will be a conflict or not. Driver behaviors, such as performance, gaze direction, and EAR, also impact the prediction result. The inputs that have no effect in each model are removed to avoid

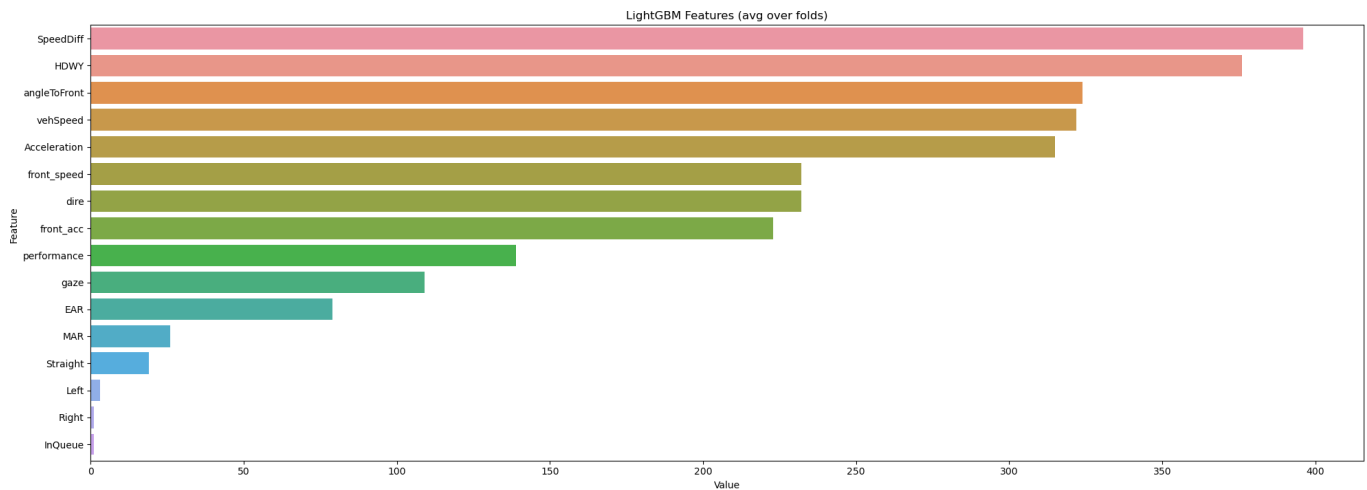

Fig. 18. Feature importance of the LGBM model.

over-fitting problems, such as safe distance and type of drivers.

Two test cases are created to further explore the influence of driver behavior information: (i) train models without driver behavior inputs and (ii) train models with driver behavior inputs. The evaluation metrics are shown in Table IV. The result shows that adding driver behavior into the model can increase model performance. Using the predicted safety indices for individual drivers, risk scores for road segments can be calculated. Table V and Table VI are the confusion matrix for predicting risk score levels in the next one or two seconds. The result shows that the model can classify the risk score levels correctly in most cases. Table VII presents more evaluation metrics of the risk score prediction results.

TABLE IV
PERFORMANCE COMPARISON BETWEEN LGBM MODELS

| Case | Evaluation Metric | TTC in next 1s | TTC in next 2s | MTTC in next 1s | MTTC in next 2s | DRAC in next 1s | DRAC in next 2s |
|---|---|---|---|---|---|---|---|
| i | Accuracy | 0.990 | 0.991 | 0.978 | 0.953 | 0.971 | 0.945 |
| ii | Accuracy | 0.999 | 0.998 | 0.990 | 0.974 | 0.978 | 0.972 |
| i | Precision | 0.992 | 0.992 | 0.977 | 0.954 | 0.970 | 0.948 |
| ii | Precision | 0.999 | 0.998 | 0.989 | 0.978 | 0.977 | 0.969 |
| i | Recall | 0.992 | 0.993 | 0.977 | 0.955 | 0.970 | 0.944 |
| ii | Recall | 0.999 | 0.998 | 0.989 | 0.977 | 0.981 | 0.972 |
| i | F1-Score | 0.993 | 0.993 | 0.977 | 0.954 | 0.971 | 0.943 |
| ii | F1-Score | 0.999 | 0.998 | 0.991 | 0.974 | 0.978 | 0.966 |

TABLE V
CONFUSION MATRIX FOR RISK SCORE LEVEL IN THE NEXT ONE SECOND

| | | Predicted Risk Score Level | | | | |
|---|---|---|---|---|---|---|
| | | Very small | Small | Medium | Large | Very large |
| Actual Risk Score Level | Very small | 9829 | 0 | 0 | 0 | 0 |
| | Small | 0 | 4706 | 928 | 1 | 0 |
| | Medium | 0 | 2 | 303 | 13 | 0 |
| | Large | 0 | 8 | 343 | 436 | 0 |
| | Very Large | 0 | 0 | 2 | 2 | 52 |

TABLE VI
CONFUSION MATRIX FOR RISK SCORE LEVEL IN THE NEXT TWO SECONDS

| | | Predicted Risk Score Level | | | | |
|---|---|---|---|---|---|---|
| | | Very small | Small | Medium | Large | Very large |
| Actual Risk Score Level | Very small | 9829 | 0 | 0 | 0 | 0 |
| | Small | 0 | 3734 | 2002 | 53 | 0 |
| | Medium | 0 | 4 | 220 | 34 | 0 |
| | Large | 0 | 7 | 295 | 379 | 1 |
| | Very Large | 0 | 1 | 5 | 9 | 52 |

TABLE VII
EVALUATION METRICS OF RISK SCORE PREDICTION WITH REAL DATA

| Case | Accuracy | Precision | Recall | F1 Score |
|---|---|---|---|---|
| Next 1 second | 0.922 | 0.831 | 0.854 | 0.779 |
| Next 2 seconds | 0.855 | 0.773 | 0.766 | 0.693 |

## VI. DRIVING SIMULATION PLATFORM

More test cases are designed in this section to validate the improved SRM model further. Considering the difficulty and hazard of collecting data for dangerous driving behaviors on the streets, such as aggressive and distracted driving. A driving simulation platform is built to gather more data.

### A. Simulation setup

To better study near misses, a driving simulation platform is assembled, as shown in Fig. 19. Logitech G29 electronic steering wheel and pedals are installed in a racing simulator cockpit. A camera is placed on top of the TV screen to capture the face of the driver. CARLA, an open-source driving simulator [39], is utilized as the driving simulation environment. Real-time traffic information is exported from CARLA and combined with drivers' behavior information. Risk scores are predicted using the improved SRM model based on the simulated data.

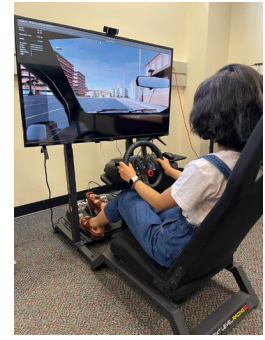

Fig. 19. Driving simulation platform.

Three test cases are designed using CARLA default maps and the maximum number of non-player characters (NPCs) allowed for each map: (i) "TOWN02", which is a tiny town with 101 vehicles and 36 pedestrians; (ii) "TOWN04", which contains a circumferential highway with 372 vehicles; and (iii) "TOWN05", which simulates a downtown area with 302 vehicles and 34 pedestrians. The CARLA simulator controls NPCs, automatically following pre-designed routes on the map. 40% of NPCs have hostile behaviors that go beyond the speed limit by 20% and change lanes more frequently. 20% of NPCs have defensive behaviors in that their maximum speed is 20% lower than the limit, and they seldom change lanes. The remaining NPCs follow the speed limit and change lanes when necessary, such as switching to a faster lane.

### B. Simulation results

Fifty-two volunteers are invited to take part in the experiments. Each volunteer is asked to drive for 20 minutes on each map. The volunteers can behave aggressively, such as tailgating, rapid lane change, speeding, or running red lights. Also, the volunteers can answer phone calls or chat with others while driving.

TABLE VIII
PERFORMANCE COMPARISON BETWEEN LGBM MODELS

| Evaluation Metric | TTC in next 1s | TTC in next 2s | MTTC in next 1s | MTTC in next 2s | DRAC in next 1s | DRAC in next 2s |
|---|---|---|---|---|---|---|
| Accuracy | 0.995 | 0.995 | 0.968 | 0.968 | 0.991 | 0.992 |
| Precision | 0.995 | 0.996 | 0.970 | 0.969 | 0.991 | 0.992 |
| Recall | 0.995 | 0.995 | 0.967 | 0.968 | 0.993 | 0.992 |
| F1-Score | 0.995 | 0.996 | 0.967 | 0.969 | 0.992 | 0.992 |

Table VIII presents the evaluation metrics of the predictions on safety indices of all drivers. Table IX shows the evaluation metrics of the final roadway risk score level prediction. Same procedure described in Section IV is applied. The results show that the improved SRM model can make accurate forecasts.

TABLE IX
EVALUATION METRICS OF RISK SCORE PREDICTION WITH SIMULATED DATA

| Case | Accuracy | Precision | Recall | F1 Score |
|---|---|---|---|---|
| Next 1 second | 0.948 | 0.805 | 0.789 | 0.794 |
| Next 2 seconds | 0.914 | 0.701 | 0.755 | 0.715 |

## VII. CONCLUSION

This paper developed a method to collect real-time traffic data from roadside and driver's behavior data inside a vehicle. The novel idea is to gather real-world data to analyze factors associated with road safety. Another contribution is that this work brings multiple data sources together, especially human factors when predicting traffic conflicts for individual drivers and calculating safety scores for road segments. With the Android APP and cloud computing, collecting data and analyzing real-time videos have become practical. The image recognition techniques are applied to detect drivers' facial expressions and vehicles on the road. Multiple driving behavior features are extracted from facial landmarks, including the EAR, MAR, focusing area, and driver's emotion. Together with roadway characteristics, vehicle speed profile, and vehicle status, the original SRM model is extended for predicting road safety. The model validation results show that the prediction is accurate. Also, conclusions can be made that road safety and drivers' risk profiles are associated with driver behaviors.

There are certain limitations that we can address in our future work. For instance, more data need to be collected to test the robustness of the model. Collecting real-world data with human drivers means we cannot intensely create hazardous traffic conditions, test drivers' reactions, and collect traffic conditions. The authors have introduced a way to simulate driving and test the SRM model. The simulation platform can be further improved to become a digital twin and be more realistic. Furthermore, other data mining methods and modeling techniques shall be introduced to strengthen the LGBM model. Correlations between the input variables shall be studied, and a hierarchical model can be built to replace the flat model in LGBM.

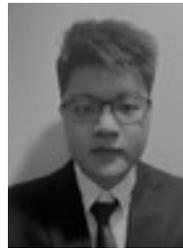
**Yufei Huang** received a B.Eng. degree in Automation from Xi'an Jiaotong University in 2016. Also, he received an M.S. degree in Systems Engineering from the University of Maryland, College Park in 2018. He is currently a Ph.D. student at Rutgers, the State University of New Jersey, studying Industrial and Systems Engineering, and a Research Assistant at the Center for Advanced Infrastructure and Transportation (CAIT). His research interests are in multi-agent systems, autonomous systems, robotics, and reinforcement learning.

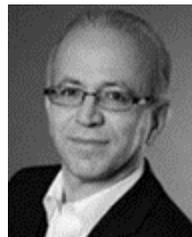
**Mohsen Jafari** (M'97) received a Ph.D. degree from Syracuse University in 1985. He has directed or co-directed a total of over 23 million U.S. dollars in funding from various government agencies, including the National Science Foundation, the Department of Energy, the Office of Naval Research, the Defense Logistics Agency, the NJ Department of Transportation,

FHWA, and industry in automation, system optimization, data modeling, information systems, and cyber risk analysis. He actively collaborates with universities and research institutes abroad. He has also been a Consultant to several Fortune 500 companies as well as local and state government agencies. He is currently a Professor and the Chair of Industrial & Systems Engineering at Rutgers University-New Brunswick. His research applications extend to manufacturing, transportation, healthcare, and energy systems. He is a member of the IIE. He received the IEEE Excellence Award in service and research.

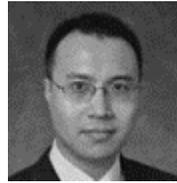

**Peter J. Jin** is an Associate Professor at Rutgers, The State University of New Jersey. He received an M.S. and Ph.D. degree from the Department of Civil and Environmental Engineering at the University of Wisconsin-Madison in 2007 and 2009, respectively. He received his B.S. degree from the Department of Automation, Tsinghua University, Beijing, China. His research interests include traffic flow theory, freeway operations, active traffic and demand management, mobile sensor data, connected vehicles, and transportation big data analytics.